\documentclass{article}
\usepackage{spconf,amsmath,graphicx}
\usepackage{amssymb,amsfonts}
\usepackage{cite}
\usepackage{booktabs}
\usepackage{url}
\usepackage{balance}
\usepackage{subfigure}
\usepackage{multirow}
\usepackage[normalem]{ulem}
\usepackage{booktabs}
\graphicspath{{./Figures/}}
\usepackage{textcomp}
\usepackage{booktabs}
\usepackage{balance}
\usepackage{pifont}
\usepackage[dvipsnames]{xcolor}

\usepackage{tikz}
\usetikzlibrary{backgrounds}
\usetikzlibrary{arrows,shapes}
\usetikzlibrary{tikzmark}
\usetikzlibrary{calc}
\usepackage{blindtext}
\usepackage{tcolorbox}
\usepackage{tikz}
\usetikzlibrary{arrows,shapes,positioning,shadows,trees,mindmap}
\usepackage[edges]{forest}
\usetikzlibrary{arrows.meta}
\colorlet{linecol}{black!75}
\usepackage{xkcdcolors} 
\usepackage{tikz}
\usetikzlibrary{backgrounds}
\usetikzlibrary{arrows,shapes}
\usetikzlibrary{tikzmark}
\usetikzlibrary{calc}
\newcommand{\highlight}[2]{\colorbox{#1!17}{$\displaystyle #2$}}

\renewcommand{\highlight}[2]{\colorbox{#1!17}{#2}}

\def\bA{{\mathbf A}}
\def\bz{{\mathbf z}}
\def\by{{\mathbf y}}

\title{Diffusion models for audio semantic communication}
\name{Eleonora Grassucci, Christian Marinoni, Andrea Rodriguez, and Danilo Comminiello \thanks{Corresponding author's email: eleonora.grassucci@uniroma1.it.}}
\address{Dept. of Information Eng., Electronics and Telecom., Sapienza University of Rome, Italy}

\begin{document}
\ninept

\maketitle

\begin{abstract}
Directly sending audio signals from a transmitter to a receiver across a noisy channel may absorb consistent bandwidth and be prone to errors when trying to recover the transmitted bits. On the contrary, the recent semantic communication approach proposes to send the semantics and then regenerate semantically consistent content at the receiver without exactly recovering the bitstream. In this paper, we propose a generative audio semantic communication framework that faces the communication problem as an inverse problem, therefore being robust to different corruptions. Our method transmits lower-dimensional representations of the audio signal and of the associated semantics to the receiver, which generates the corresponding signal with a particular focus on its meaning (i.e., the semantics) thanks to the conditional diffusion model at its core. During the generation process, the diffusion model restores the received information from multiple degradations at the same time including corruption noise and missing parts caused by the transmission over the noisy channel. We show that our framework outperforms competitors in a real-world scenario and with different channel conditions. Visit the project page to listen to samples and access the code: \url{https://ispamm.github.io/diffusion-audio-semantic-communication/}.
\end{abstract}

\keywords Audio Restoration, Generative Semantic Communication, Audio Inverse Problems, Diffusion Models

\section{Introduction}



Audio communication is the task of transmitting an audio signal from a sender over a noisy channel that can degrade and corrupt the information up to a receiver that should then retrieve the received content. However, sending the complete signal may absorb considerable bandwidth and recovering the complete bitstream at the receiver may be error-prone. This has always been considered a very tight constraint in wireless communications.

Recently, with the upcoming rise of 6G communications, semantic communication frameworks have replaced classical wireless systems. The promising aspects of semantic communication lie in the ability to regenerate content preserving the meaning of the transmission (i.e., the semantics) without necessarily recovery the exact bit sequence \cite{Dai2021CommunicationBT, Barbarossa2023COMMAG}. In recent years, few audio semantic communication frameworks have been proposed, encoding the semantics of speech signals with the help of neural networks \cite{Weng2021ICC, Xiao2023ICASSP, Han2022SemanticPreservedCS}. Concurrently, generative models have been demonstrated to be powerful and robust tools to enhance semantic communication frameworks \cite{grassucci2023enhancing} due to their ability to generate content from the received semantic information \cite{Lee2023EnergyEfficientDS}, even when extremely degraded and corrupted \cite{Grassucci2023GenerativeSC}.

In this paper, we address the problem of audio communication over a noisy channel, formulating it as an inverse problem in which the transmission deteriorates and corrupts data while the model tries to restore the original audio or its semantic aspects. Under this formulation, the problem moves to solving an audio inverse problem in which diffusion models excel \cite{Moliner2023ICASSP, Peer2023ICASSP, Richter2023TASLP}.

 To do so, we define a novel audio semantic communication framework, whose core is a latent diffusion model conditioned on textual semantics to enhance the generation results. The sender transmits lower-dimensional latent representations of the audio and of its caption to the receiver. The latter solves the inverse problem by restoring the audio from the channel noise and inpainting the missing parts that have been lost in the transmission over the channel. This is done by leveraging the range-null space decomposition that ensures consistency with the inverse problem formulation and realness according to the data distribution \cite{Wang2022ZeroShotIR}. While doing this, the diffusion model leverages the textual semantic information to ensure semantically consistent outputs and improve the quality of generation.
We conduct an experimental evaluation on a real-world dataset and we show that the proposed framework is able to denoise speeches and real-world sounds or audio scenes in the case of heavily corrupted received information. Moreover, the proposed framework inpaints meaningful speeches and sounds in audio clips with missing parts although the received semantic information may be corrupted by the noisy channel.

\begin{figure}[t]
    \centering
    \includegraphics[width=\linewidth]{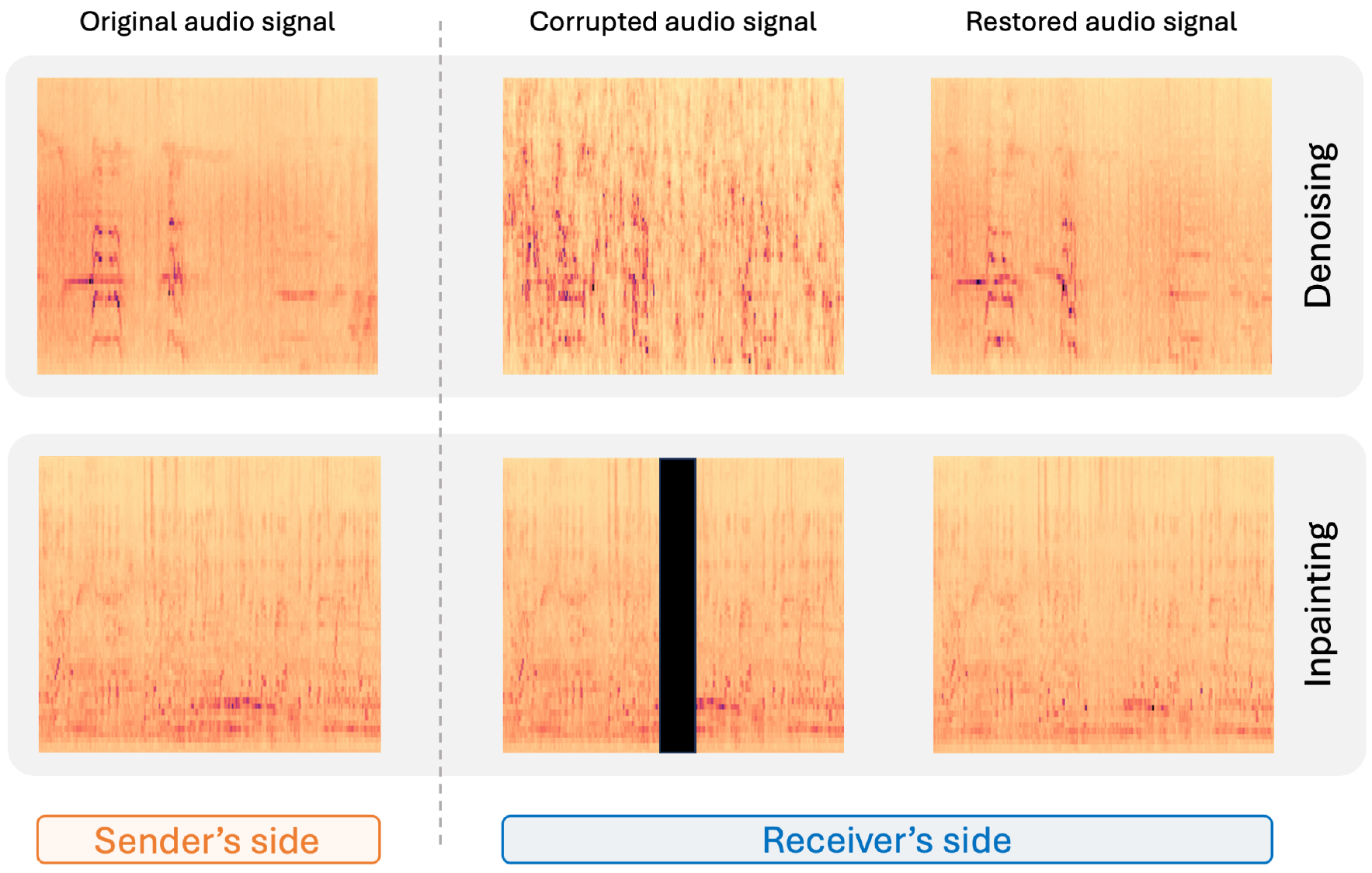}
    \caption{Results of the proposed framework on the denoising and inpainting tasks performed on low-dimensional representations of audio signals and semantics corrupted by a communication channel.}
    \label{fig:res_intro}
\end{figure}

Summarizing our contributions: i) To the best of our knowledge, we propose the first diffusion model-based framework for audio semantic communication; ii) We design a reverse sampling procedure to perform multiple restorations at the same time, such as denoising and inpainting even in the case of highly degraded channel conditions; iii) We show the effectiveness of the proposed framework in real-world scenarios, including both speeches and sounds proving its superiority with respect to state-of-the-art comparisons.

The rest of the paper is organized as follows: in Sec.~\ref{sec:method} we formulate the problem setting and derive the proposed framework. Section~\ref{sec:exp} shows the experimental evidence of the proposed method, while we draw conclusions in Sec.~\ref{sec:con}.


\section{Audio semantic communication}
\label{sec:method}

\subsection{Problem Formulation}
\label{subsec:problem}

Real-world communication systems face physical challenges due to the communication channel that may distort, corrupt, and lose portions of the transmitted signal. A quantitative way to characterize the amount of noise added to the transmitted content due to the channel conditions is the PSNR, as:

\begin{equation}
    \text{PSNR} = 10 \log \frac{P}{\sigma^2_c},
\end{equation}

\noindent where $P$ is the signal power and $\sigma^2_c$ the channel variance. Lower values of the PSNR represent bad channel conditions and potential heavy data corruption, while high values of the PSNR stand for good transmissions.
In addition to the corruption due to the noise, there may be missing portions of the received content due to losses in the case of bad channel conditions. In this scenario, the receiver should be able to fill the gap with semantically-consistent content.

Therefore, we can formulate the transmission of a content $\mathbf{z}$ over the channel as $\mathbf{y} = \mathbf{A}\mathbf{z} + \mathbf{n}$, where $\mathbf{n} \sim \mathcal{N}(0, \sigma^2_c\mathbf{I})$ is the noise added by the channel, and $\mathbf{A}$ the matrix of the corruption, indicating the missing portions of the transmitted content. The received content $\mathbf{y}$ is therefore a noisy and masked version of the original transmitted content $\mathbf{z}$. Consequently, we can handle such communication formulation as an inverse problem and try to solve it with diffusion models.

\subsection{Audio Semantic Communication Framework}

\begin{figure}
    \centering
    \includegraphics[width=\linewidth]{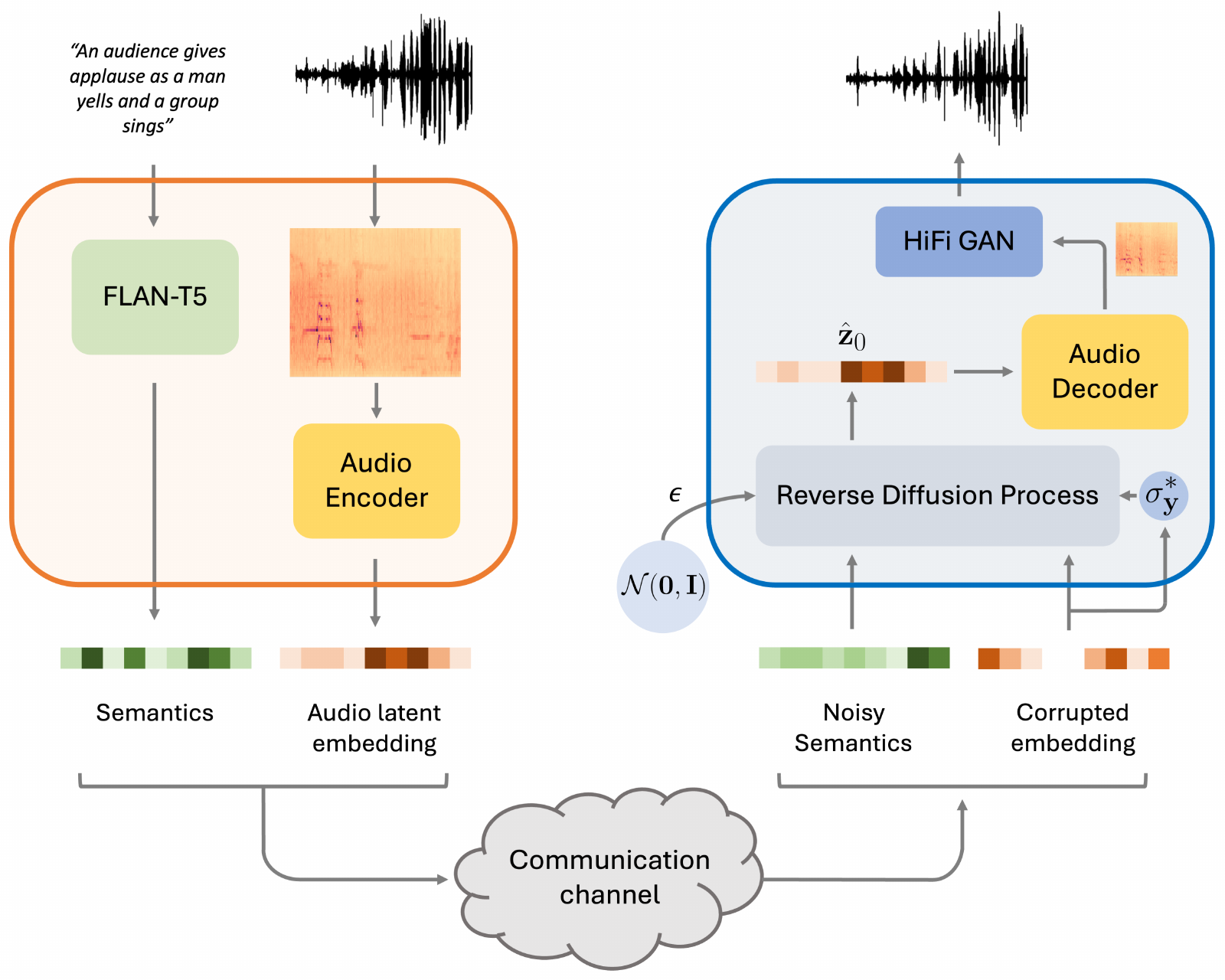}
    \caption{Proposed diffusion model for audio semantic communication framework. The original audio and the corresponding semantics are encoded and transmitted over the channel. The receiver restores the audio according to its semantics.}
    \label{fig:framework}
\end{figure}

We develop the proposed audio semantic communication framework on top of a text-to-audio latent diffusion model \cite{Ghosal2023TexttoAudioGU}. At the sender side, we first extract the mel-spectrogram of the audio waveform, then we encode it by means of the VAE encoder into the latent space. Simultaneously, the text encoder extracts the latent representation of the audio caption. The two lower-dimensional representations are transmitted over the communication channel to the receiver.
Figure \ref{fig:framework} shows the proposed framework.

\textbf{Audio Encoder and Vocoder.} The VAE encoder squeezes the mel-spectrogram in a lower-dimensional latent representation $\mathbf{z}_0 \in \mathbb{R}^{C\times L/r \times F/r}$, in which $C$ are the channels, $L$ the time, $F$ the frequency and $r$ the compression. Involving the pretrained VAE \cite{Liu2023AudioLDMTG}, we consider the best setting of $C=8$ and $r=4$ in which the residual U-Net blocks of the encoder-decoder structure have been trained to maximize the evidence lower bound while minimizing the adversarial loss.

\textbf{Textual Encoder.} To encode textual captions into latent representations that can be transmitted over the channel and then leveraged by the diffusion model, we involve the pretrained LLM FLAN-T5-Large \cite{Chung2022ScalingIL}, following \cite{Ghosal2023TexttoAudioGU}. FLAN-T5 has 780M parameters and it has been trained on a large-scale chain-of-thought (CoT) and instruction-based dataset.

\textbf{Latent Diffusion Model.} The core of our framework is a latent diffusion model \cite{Rombach2021latent} that receives the corrupted information and solves the inverse problems. The forward diffusion is a Gaussian Markov chain that destroys the data distribution into a standard Gaussian distribution in $T$ steps with predefined noise schedule $0 < \beta_1 < ... < \beta_T < 1$, following the transition probabilities:

\begin{equation}
\begin{split}
\label{eq:forward}
    q(\mathbf{z}_t | \mathbf{z}_{t-1}) = \mathcal{N}(\sqrt{1-\beta_t}\mathbf{z}_{t-1}, \beta_t \mathbf{I}), \\
    q(\mathbf{z}_t | \mathbf{z}_0) = \mathcal{N}(\sqrt{\bar{\alpha}_t}\mathbf{z}_0, (1-\bar{\alpha}_t) \mathbf{I}).    
\end{split}
\end{equation}

\noindent Note that $\alpha_t = 1- \beta_t$ and that $\bar{\alpha}_t = \prod_{i=1}^t \alpha_t$. The process output is $\mathbf{z}_T \sim \mathcal{N}(\mathbf{0}, \mathbf{I})$.

The underlying model is the U-Net backbone of StableDiffusion, having four encoder blocks, a middle block and four decoder blocks.

We involve the TANGO pretrained latent diffusion model \cite{Ghosal2023TexttoAudioGU} and replace the standard conditional reverse process with a process able to solve inverse problems. We describe it in the next subsection.

\subsection{Solving Audio Semantic Communication Inverse Problems}

Recall that from the range-null space decomposition, given a linear operator $\mathbf{A}$, any sample $\mathbf{z}$ can be decomposed into two parts: i) the \highlight{orange}{range space of $\mathbf{A}$}, and ii) the \highlight{blue}{null space of $\mathbf{A}$} \cite{Schwab_2019, Wang2022arxiv}. Therefore, the sample $\mathbf{z}$ can be written by

\begin{equation}
\label{eq:sample}
    \mathbf{z} = \highlight{orange}{$\mathbf{A}^\dagger \mathbf{A}\mathbf{z}$} + \highlight{blue}{$(\mathbf{I} - \mathbf{A}^\dagger \mathbf{A}) \mathbf{z}$}.
\end{equation}

Let us consider a generic inverse problem formulation $\mathbf{y} = \mathbf{A}\mathbf{z}$, the solution to this problem consists in an audio latent vector $\hat{\mathbf{z}}$ that satisfies the two constraints:
\begin{align}
    \highlight{LimeGreen}{$\mathbf{A}\hat{\bz} = \mathbf{y}$} \\
    \hat{\mathbf{z}} \sim q(\mathbf{z}),
\end{align}

\noindent that are, respectively, \highlight{LimeGreen}{consistency} and \textbf{realness}. Recalling the sample formulation in \eqref{eq:sample} and applying the operator $\mathbf{A}$, the range space becomes $\mathbf{y}$, while the null space becomes $\mathbf{0}$ since $\bA \bz = \bA \highlight{orange}{$\bA^\dagger \bA \bz$} + \bA \highlight{blue}{$(\mathbf{I} - \bA^\dagger \bA) \bz$} = \bA \bz + \mathbf{0} = \by$. Therefore, for any inverse problem of this form, we can formally build the solution $\hat{\bz} = \highlight{orange}{$\bA^\dagger\by$} + \highlight{blue}{$(\mathbf{I}-\bA^\dagger \bA) \tilde{\bz}$} $ that satisfies the consistency constraint, whatever $\tilde{\bz}$ is. The solution $\tilde{\bz}$, however, determines whether the solution satisfies the realness constraint too. The scope of the training is therefore finding the $\tilde{\bz}$ such as $\hat{\bz} \sim q(\bz)$, and the diffusion model can be trained to generate the proper null space \highlight{blue}{$(\mathbf{I}-\bA^\dagger \bA) \tilde{\bz}$} for the range space \highlight{orange}{$\bA^\dagger\by$}.

However, intermediate states $\bz_t$ of the reverse process are noisy and this can break the harmony between the range and the null space \cite{Wang2022ZeroShotIR}. To avoid this misalignment, the mean and the variance of the intermediate state $p(\bz_{t-1}|\bz_t, \bz_0)$ can be reparameterized to arrive at the desired output $\bz_0 \sim q(\bz)$ as

\begin{align}
    \mu_t(\bz_t, \bz_0) = \frac{\sqrt{\bar{\alpha}_{t-1}}\beta_t}{1-\bar{\alpha}_t}\bz_0 + \frac{\sqrt{\alpha_t}(1-\bar{\alpha}_{t-1})}{1-\bar{\alpha}_t} \bz_t \\
    \sigma^2_t = \frac{1-\bar{\alpha}_{t-1}}{1-\bar{\alpha}_t}\beta_t.
\end{align}

\noindent We can reverse \eqref{eq:forward} to estimate $\bz_0$ from $\bz_t$ and from the predicted noise $\mathbf{\epsilon}_t = \mathcal{Z}_\theta(\bz_t, t)$ and formulate the estimated $\bz_0$ as

\begin{equation}
    \bz_{0|t} = \frac{1}{\sqrt{\bar{\alpha}_t}} \left( \bz_t - \mathcal{Z}_\theta \right).
\end{equation}

\noindent Finally, the estimated $\bz_{0|t}$ is computed by

\begin{equation}
\label{eq:z0t}
    \hat{\bz}_{0|t} = \highlight{orange}{$\bA^\dagger \by$} + \highlight{blue}{$(\mathbf{I} - \bA^\dagger \bA) \bz_{0|t}$} = \bz_{0|t} - \highlight{LimeGreen}{$\bA^\dagger (\bA \bz_{0|t}- \bA\bz)$}.
\end{equation}

However, in the case of noisy inverse problems as formulated in Subsection \ref{subsec:problem} for communications, a further noisy term $\bA^\dagger \mathbf{n}$ would be introduced in \eqref{eq:z0t}, producing final noisy samples. Therefore, we can introduce two parameters in the reverse process to adapt the formulation to noisy inputs:

\begin{align}
    \hat{\bz}_{0|t} = \bz_{0|t} - \Sigma_t \highlight{LimeGreen}{$\bA^\dagger (\bA \bz_{0|t} - \by)$}, \\
    \hat{p}(\bz_{t-1}|\bz_t, \hat{\bz}_{0|t}) = \mathcal{N}(\mu_t(\bz_t, \hat{\bz}_{0|t}), \Phi_t\mathbf{I}),
\end{align}

\noindent in which $\Sigma_t$ scales the range space correction \highlight{LimeGreen}{$\bA^\dagger (\bA \bz_{0|t}- \by)$} and $\Phi_t$ scales the noise $\sigma_t \mathbf{\epsilon}$ in $p(\bz_{t-1}|\bz_t, \hat{\bz}_{0|t})$. The two terms need to satisfy some constraints: i) $\Sigma_t$ has to tend to the identity matrix so as to maximize the consistency through the range space correction \highlight{LimeGreen}{$\bA^\dagger (\bA \bz_{0|t} - \by)$}, while ii) $\Phi_t$ has to guarantee that the noise variance in $\bz_{t-1}$ is equal to $\sigma_t^2$ so that it can be removed by the pretrained model that estimates the noise through $\mathcal{Z}_\theta$. We can approximate $\bA^\dagger \mathbf{n} \sim \mathcal{N}(\mathbf{0}, \sigma^2_\by \mathbf{I})$, where $\sigma^2_\by$ is the variance of the noise in the received latent vector $\by$, which is the variance $\sigma^2_c$ of the channel noise rescaled according to the original data range. With this approximation, we can simplify $\Sigma_t = \lambda_t \mathbf{I}$ and $\Phi_t = \gamma_t \mathbf{I}$ \cite{Wang2022ZeroShotIR}. Given that the intermediate state $\bz_{t-1}$ is equal to

\begin{equation}
    \bz_{t-1} = \frac{\sqrt{\bar{\alpha}_{t-1}}\beta_t}{1-\bar{\alpha}_t}\hat{\bz}_{0|t} + \frac{\sqrt{\alpha_t} (1-\bar{\alpha}_{t-1})}{1-\bar{\alpha}_t}\bz_t + \sigma_t\epsilon,
\end{equation}

\noindent we can satisfy constraint i) by setting:

\begin{equation}
    \gamma_t = \sigma_t^2 - (\frac{\sqrt{\bar{\alpha}_{t-1}}\beta_t}{1-\bar{\alpha}_t} \lambda_t \sigma_\by)^2,
\end{equation}
\begin{equation}
    \lambda_t = \begin{cases}
        1, \qquad \sigma_t \ge \frac{\sqrt{\bar{\alpha}_{t-1}}\beta_t}{1-\bar{\alpha}_t}\sigma_\by \\
        \sigma_t/ \sigma_\by, \qquad \sigma_t < \frac{\sqrt{\bar{\alpha}_{t-1}}\beta_t}{1-\bar{\alpha}_t}\sigma_\by
    \end{cases}
\end{equation}

\noindent and constraint ii) with:

\begin{equation}
    \left(\frac{\sqrt{\bar{\alpha}_{t-1}}\beta_t}{1-\bar{\alpha}_t} \lambda_t \sigma_\by \right)^2 +\gamma_t = \sigma_t^2.
\end{equation}

\noindent With the above formulation, the only parameter that has to be set is $\sigma_\by$, from which the denoising ability of the sampling procedure depends. We discuss it in the next Subsection.

\subsection{Automating the choice of denoising hyperparameter}
\label{automated-sigmay}
The optimal value $\sigma^*_\mathbf{y}$  of $\sigma_\mathbf{y}$ strictly depends on the variance of the noisy observation $\mathbf{y} \sim \mathcal{N}(0, \sigma^2_\mathbf{y})$. A manual setting of such a hyperparameter, as suggested in previous works \cite{Wang2022ZeroShotIR}, is unfeasible in a communication scenario where the receiver is unaware of the channel conditions and of the distortions they may have applied to the transmitted content. In addition, we notice that $\sigma^*_\mathbf{y}$ directly hinges on the data range too since the standard deviation is scaled as the data distribution scales.
Therefore, we propose to automatically compute the optimal value $\sigma^*_\mathbf{y}$ that adaptive changes depending on the range and on the standard deviation of the received data $\mathbf{y}$ following:

\begin{equation}
\label{eq:sigmay}
    \sigma^*_\mathbf{y} = \left(\max \left(\mathbf{y} \right)-\min \left(\mathbf{y} \right) \right) \cdot \sigma_\mathbf{y}.
\end{equation}

\noindent Equipped with this formulation, the proposed method at the receiver side can automatically compute the optimal value for denoising without requiring any human feedback or knowledge about the channel conditions. This transforms the proposed method in an end-to-end method robust to different and unknown channel conditions.

\section{Experimental Evaluation}
\label{sec:exp}

We perform the experimental evaluation on AudioCaps \cite{audiocaps}, a real-world large-scale dataset of about $46$k audio clips with human-collected text pairs starting from the AudioSet dataset \cite{AudioSet}. We resample all samples to $16$ kHz and standardize the length to be $10$ seconds long. We perform two sets of experiments, in a denoising-only scenario and with the inpainting task, both under different channel conditions with PSNR values in the set $[15, 17.5, 20, 30]$. 


\begin{table}
\label{tab:results-denoising}
\caption{Denoising results according to SNR and FAD. The proposed framework better denoises received samples according to both metrics in every test we conduct.}
\centering
\resizebox{\linewidth}{!}{
\begin{tabular}{l|cc|cc|cc|cc}
\toprule
 PSNR$\rightarrow$ &
  \multicolumn{2}{c|}{15} &
  \multicolumn{2}{c|}{17.5} &
  \multicolumn{2}{c|}{20} &
  \multicolumn{2}{c}{30} \\ 
                           Model & SNR$\uparrow$     & FAD$\downarrow$     & SNR$\uparrow$     & FAD$\downarrow$     & SNR$\uparrow$     & FAD$\downarrow$     & SNR$\uparrow$     & FAD$\downarrow$     \\ \midrule
\multicolumn{1}{l|}{N2N}  & -8.08 & 22.07 & -6.81 & 20.42 & -5.16 & 18.25 & \textbf{1.74}  & 11.04 \\
\multicolumn{1}{l|}{Ours} & \textbf{-2.88} & \textbf{21.24} & \textbf{-2.63} & \textbf{10.87} & \textbf{-2.74} & \textbf{8.38}  & -2.57 & \textbf{3.75}  \\ \bottomrule
\end{tabular}}
\end{table}

\subsection{Denoising}
The denoising scenario faces the case in which both the low-dimensional latent representation of the audio signal and the semantics are heavily affected by the noise coming from the communication channel.
To mimic the behavior of the channel, we apply Gaussian noise that adheres to the predefined PSNR values.
When using our method, we set $\sigma^*_\mathbf{y}$ as defined in Section~\ref{automated-sigmay}, resulting in higher values for lower PSNRs. For instance, $\sigma_\by^*$ is, on average, equal to $68$ when the PSNR equals $15$. Moreover, we employ $1k$ steps in the diffusion process and a guidance scale equal to $3$.
We compare our solution with a U-Net-based approach that extends Noise2Noise (N2N) to the speech denoising task \cite{kashyap21_interspeech}. In particular, we take the original architecture and, to simulate a deteriorated channel, we retrain it on a noisy version of the AudioCaps data set till convergence. Compared to our approach, this method has two principal differences. First, N2N operates directly on the input data, while our model forges on the lower-dimensional latent space and crucially has, therefore, lesser bandwidth requirements. Second, N2N does not employ the semantic information provided by captions to guide the generation process, as ours does instead.


We evaluate the approaches with two metrics, the Signal-to-Noise Ratio (SNR) and the Fréchet Audio Distance (FAD) \cite{kilgour2019frechet}, in the four levels of PSNR. The SNR quantifies the ratio of the power of the desired signal to the strength of the unwanted noise. A higher SNR value indicates a better-denoised audio signal. However, reducing noise is only one side of the medal, as the denoising process can introduce distortions. To account for this fact, we also consider FAD, a reference-free metric that correlates more closely with human perception.
As shown in Tab.~\ref{tab:results-denoising}, our approach provides the best results, both in terms of SNR and FAD, accounting for the semantics of the audio samples. Moreover, our method leads to lower band occupancy. An example of the result of the denoising (first row) and then inpainting (second row) tasks is depicted in Fig.~\ref{fig:res_intro}.



\begin{table}[t]
\centering
\caption{Inpainting results as measured by FAD metrics (the lower the better) on the whole audio (All$\downarrow$) and on the inpainted part only (Inp$\downarrow$). Our method provides the best results, especially in the case of bad channel conditions. Average over multiple runs.}
\label{tab:inp_results}
\resizebox{\linewidth}{!}{
\begin{tabular}{@{}l|cc|cc|cc|cc@{}}
\toprule
 PSNR$\rightarrow$           & \multicolumn{2}{c|}{15}    & \multicolumn{2}{c|}{17.5}  & \multicolumn{2}{c|}{20}    & \multicolumn{2}{c}{30}    \\ 
 Model & All$\downarrow$ & Inp$\downarrow$ & All$\downarrow$ & Inp$\downarrow$ & All$\downarrow$ & Inp$\downarrow$ & All$\downarrow$ & Inp$\downarrow$ \\ \midrule

AudioLDM \cite{Liu2023AudioLDMTG}        & 2.23 & 14.89 & 2.25 & 14.13 & 2.29 & 13.95 & 2.32 & 12.11 \\
Repaint \cite{Lugmayr2022RePaintIU} & 4.98 & 21.83 &  3.02 & 19.84 &  2.95 & 16.21 & 2.44 & 15.01  \\
Ours            & \textbf{2.14} & \textbf{11.95} & \textbf{2.16} & \textbf{12.52} & \textbf{1.98} & \textbf{10.37} & \textbf{2.08} & \textbf{10.33}  \\ \bottomrule
\end{tabular}}
\end{table}

\subsection{Inpainting}
Another scenario we can encounter during transmission through a communication channel is losing part of the information. In this scenario, a receiver equipped with the proposed method can regenerate the missing content in a semantically consistent way. More formally, the receiver obtains the latent representation of the audio signal with a missing portion to retrieve by solving the related inverse problem. The sender also transmits the corresponding semantics, subject to channel noise, that the receiver can leverage to guide the generation process.
To reproduce the behavior of the channel, we apply additive white gaussian noise (AWGN) according to the chosen PSNR values to the semantics, and we mask a 1-second-long section of the audio latent representation to simulate a loss of information.
While $\sigma_\by$ should be equal to zero when dealing with non-noisy inverse problems, here we set $\sigma_\by$ according to \eqref{eq:sigmay} computed on the caption embeddings. Indeed, we notice that noisy conditioning can, in turn, introduce unknown noise in the reverse diffusion process, producing dirty samples. Therefore, we propose to treat this task as a (slightly) noisy inverse problem as well and jointly perform denoising while inpainting the missing part.

We compare our model with two state-of-the-art approaches: Tango\cite{Ghosal2023TexttoAudioGU} + RePaint\cite{Lugmayr2022RePaintIU} and AudioLDM \cite{Liu2023AudioLDMTG}. The first consists of replacing the reverse diffusion process of TANGO with one inspired by RePaint \cite{Lugmayr2022RePaintIU}, meaning that this method shares the same architecture as ours but presents a different sampling procedure.
The second comparative method is AudioLDM, a text-to-audio system designed to learn continuous audio representations from CLAP \cite{laionclap2023} embeddings and capable of performing zero-shot audio inpainting and style transfer. In this case, the architecture is different from ours, thus making it possible to assess the performance of a distinct framework on this task. Since it is not performed natively by AudioLDM, we apply noise to the embeddings used for conditioning the sampling process following the same four PSNR levels.

\begin{figure}
    \centering
    \includegraphics[width=\linewidth]{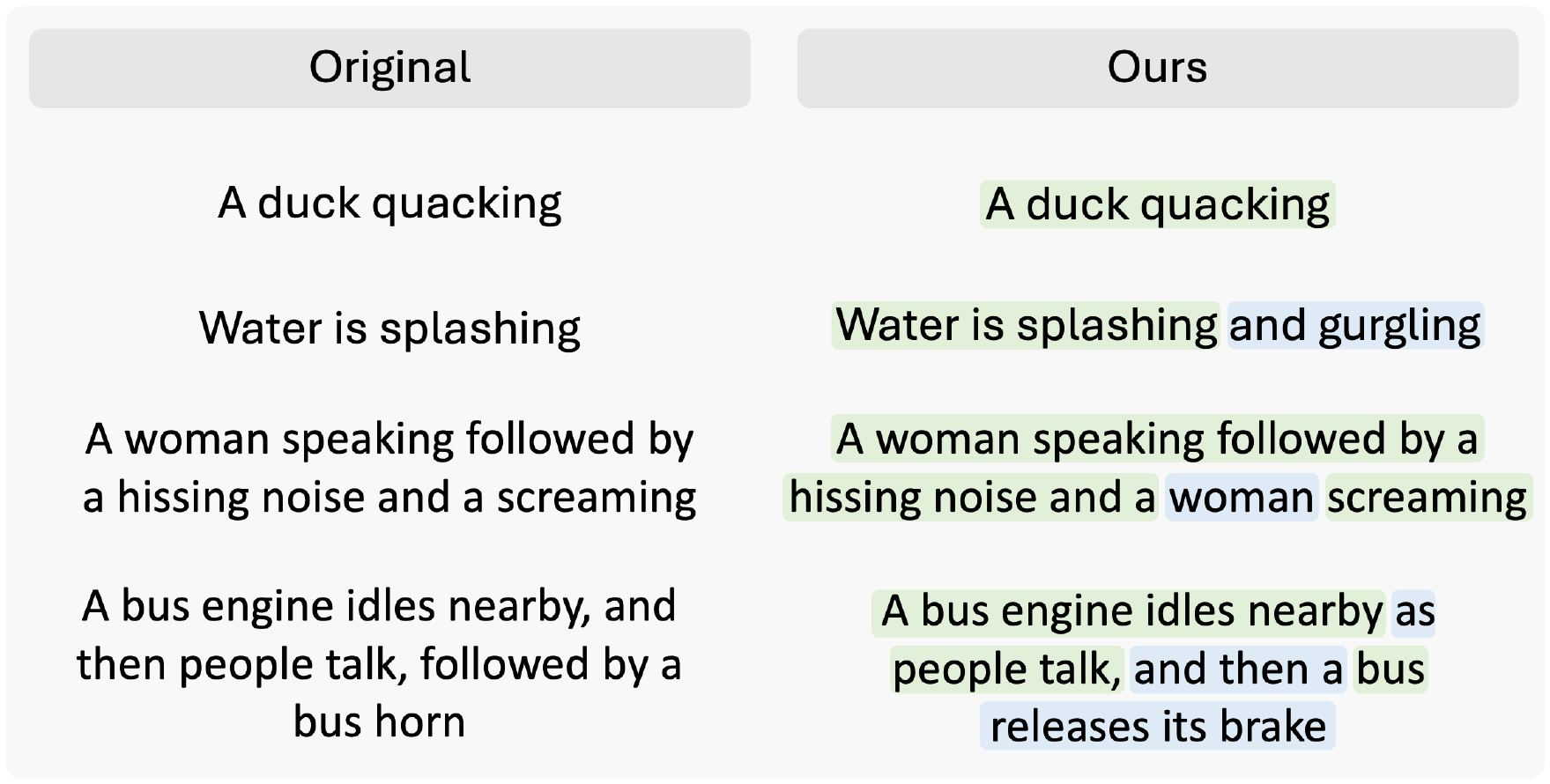}
    \caption{Captions generated by applying the Whisper Audio Captioning model. The left column shows captions on clean audio samples. On the right, the captions are derived from the same captioning model applied to audio samples generated with our approach. }
    \label{fig:audio_captions}
\end{figure}

We evaluate the three approaches with the Fréchet Audio Distance (FAD) on the entire duration of the audio sample (10 seconds) and on the masked section only (1 second). We refer to these as \textit{All} and \textit{Inp} FAD. Indeed, we notice that calculating the metrics focusing on the inpainted part of the audio allows a better estimation of the effectiveness of the methods analysed on the inpainting task. Complementary, the \textit{All} FAD takes into account any distortions introduced by the models on known parts.
Table~\ref{tab:inp_results} reports FAD values associated with the corresponding four selected PSNR values. Our method achieves the best results in all the most challenging configurations, still being able to compete with the state of the art for higher PSNR values. Moreover, we perform a semantic evaluation of the inpainted audio. We apply Whisper Audio Captioning V2 \cite{kadlvcik2023whisper} to generate captions for audio samples generated with our model (with PSNR=20) and analyse its impact on the semantics. We repeat this process to produce the captions associated with the original sound, thus enabling a fair comparison between our samples and the unmasked ones. Figure~\ref{fig:audio_captions} shows random captions of original vs. inpainted audio by our method and highlights the consistency of our results. Indeed, our method produces audio with congruous captions with respect to original uncorrupted audio, thus proving that the proposed framework preserves semantics in restored samples.



\section{Conclusion}
\label{sec:con}
In this paper, we present a novel generative audio semantic communication framework that addresses the problem of denoising or inpainting the lower-dimensional latent representation of audio samples with the help of semantics. Our solution, which provides better results on the metrics considered, has two remarkable features: (1) it allows efficient use of the communication channel thanks to a reduced amount of information the sender needs to transmit to the receiver; (2) it allows efficient estimation of the original transmitted data by exploiting semantics, even when the channel suffers from high noise or when part of the content is lost. 

\balance
\bibliographystyle{IEEEbib}
\bibliography{Biblio}

\end{document}